\documentclass[%
 reprint,
 amsmath,amssymb,
 aps,
]{revtex4-1}

\usepackage{xcolor}

\usepackage{graphicx}
\usepackage{dcolumn}
\usepackage{bm}
\usepackage{texshade}

\begin{document}


\title{Phase diagram of two-patch colloids with competing anisotropic and isotropic interactions}

\author{Isha Malhotra and Sujin B. Babu}
 \altaffiliation[Also at ]{Department of physics, Indian Institute of Technology, Delhi-110016}
 
\date{\today}

\begin{abstract}
Patchy particles are considered to be a good model for protein aggregation. We calculate the equilibrium phase diagram of two-patch colloidal particles undergoing aggregation along with isotropic potential. This investigation demonstrates the coexistence of different phases like disordered clusters, chains, crystals and bundles depending on the relative strength of isotropic and anisotropic potential, which are also reported for glucose isomerase proteins. We also show that the formation of network of bundles is metastable against the formation of thermodynamically favored finite sized bundles which are also metastable along with crystals. These bundles appear to be helical in structure similar to that observed in sickle cell hemoglobin.
\end{abstract}

\pacs{Valid PACS appear here}
\maketitle


Human diseases such as cataract, sickle cell anaemia, Alzheimer's disease, Amyotrophic lateral sclerosis are caused due to undesired protein aggregation\cite{siezen1985opacification,vekilov2007sickle,ghiso2002amyloidosis,patel2015liquid}. However, it is also desired to build protein crystals via self assembly to get insight into the structure of proteins from a fundamental point of view \cite{berman2006worldwide} and for various pharmaceutical applications like targeted drug delivery \cite{basu2004protein}. Patchy particles are considered to be a very good model for protein aggregation \cite{guo2014predictive}, they can also be used in imitating atomic orbitals \cite{wang2012colloids}, tailor made colloidal structures via self assembly \cite{manoharan2003dense,zhang2004self} etc. It has already been shown that the anisotropic potential developed by Kern and Frenkel \cite{kern2003fluid} called patchy particles can mimic certain type of equilibrium protein crystals \cite{roberts2014role}. Later patchy particles along with isotropic interaction \cite{liu2009self,vekilov2005two,liu2007vapor,gliko2005metastable,gliko2007metastable} were found to be closer to experimental results than just isotropic or anisotropic models. It was shown that by playing with the number of patches along with short range isotropic interaction reproduced the liquid-liquid coexistence curves of gamma-crystalline and lyzozyme \cite{liu2007vapor} proteins. The lyzoyme protein shows a transition from chains to bundles \cite{woodard2014gel} as seen in amyloid fibers believed to be responsible for some of the neuro-degenerative diseases. Bundle formation is also observed in other biological systems like the actin filaments \cite{sept2001thermodynamics}. Chain to bundle transition was shown for a two patch particle where irreversible patchy interaction was coupled with reversible isotropic interaction \cite{malhotra2018aggregation}. Huisman \textit{et al.}\cite{huisman2008phase} also observed the formation of bundles and they showed that transition is similar to sublimation transition of polymers. Preisler \textit{et al.}\cite{preisler2013phase} also observed the formation of a metastable tubes for the case of particles with a single patch.

In the present work we have simulated two patch colloidal particles interacting via reversible anisotropic and isotropic potential. Tuning both anisotropic and isotropic interaction we are able to observe the existence of a rich phase diagram consisting of phases such as chains + crystals, thermodynamically favored bundles +crystals, disordered clusters + crystals and disordered clusters.  Whitelam \textit{et al.}\cite{Whitelam} proposed a 2D model where he showed that different pathways for crystallization is possible by tuning the relative strength of the nonspecific and specific interactions. For $2$ patch particles  Li \textit{et al.}\cite{li} observed the formation of different stable crystal structures when the patch well depth was kept twice of isotropic well depth, but they were not able to observe fiber-bundle transition. Audus \textit{et al.} \cite{Audus} also studied the phase behavior for five patch particle with competing isotropic and anisotrpic interaction, where they observed only liquid-liquid phase separation.

\begin{figure}
\includegraphics[height=2.5cm,width=8.5cm]{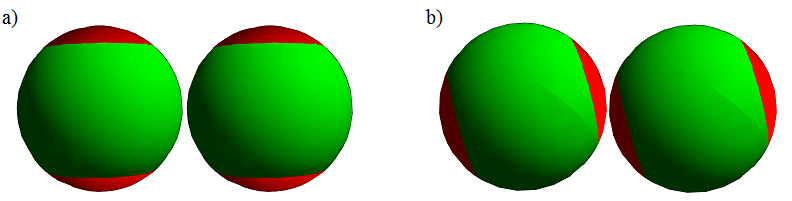}
  \caption{(a). Particles are in interaction range and forming $NPI$ bond with a probability $1-exp(-u_i)$. (b) Particles are in interaction range and forming $P$ bond with a probability $1-exp(-u_a)$. }
    \label{PNPI}
    \end{figure}

Our model consists of hard spheres of diameter $\sigma$, each particle is associated with a unit patch vector 
$\hat{\bf v}_{i}$ which defines the two oppositely located patches. We have isotropic potential coupled with Kern and Frenkel \cite{kern2003fluid} anisotropic potential given by 
\begin{equation}
 U (\mathbf{r}_{i,j},  {\bf v}_{i} , {\bf v}_{j})  = 
 \begin{cases}
   \infty\hspace{2.2cm} r_{i,j}\leq \sigma 
   \\
   -(u_{i} + u_{a})\quad \sigma < r_{i,j}\leq\sigma(1+\epsilon)
\\
0\hspace{2.3cm}  r_{i,j}>\sigma(1+\epsilon)
\end{cases}
\label{e.1}
 \end{equation}

 where $r_{i,j}$ is the distance of centre of mass of the $i_{th}$ particle to the $j_{th}$ particle and $\epsilon$ is the interaction range. $u_i$ and $u_a$ are the depths of square well for isotropic and anisotropic interactions respectively.  $u_a$ is non zero only if $ {\textbf{\^r}}_{i,j}.{\textbf{\^v}}_{i}\ > \cos \omega \text{and}\         {\textbf{\^r}}_{j,i}.{\textbf{\^v}}_{j}>\cos \omega$, where $\omega$ is the semi solid angle of a cone having vertex at the center of a sphere which defines the patch size. In the present work we have used $\sigma=1$, $\epsilon=0.1$ and $\omega=\pi/8$ which ensures that only one bond per patch is possible \cite{malhotra2018aggregation}. In the present study, all quantities are reported in reduced units, hence bond strength is measured in units of $u/k_BT$ where $k_BT=1$. Reduced time is given by $t/t_0$, where $t_0$ is the time taken by a particle to travel it's own diameter \cite{malhotra2018aggregation} and the volume fraction is defined as $\phi=\pi/6 N_{tot}/L^3$.  We have kept $\phi=0.02$ which corresponds to a total number of $N_{tot}=4774$ particles. These particles are randomly distributed in a cubic box of fixed length $L=50$ with periodic boundary condition. Every simulation step is divided into two steps, first one is the movement step, where we randomly select $2 N_{tot}$ particles which are either rotated with a step size $s_R$ or translated with a step size $s_T$ \cite{prabhu2014brownian}. This ensures that every particle undergoes translational and rotational diffusion in an uncorrelated and independent manner in every step. We have fixed $s_T=0.013$ and $s_R=0.018$ to ensure the correct diffusional behavior \cite{prabhu2014brownian} . All movement steps which lead to the overlap of spheres or breakage of bonds are rejected. Second step is the cluster construction step, when two monomers are in interaction range and their patches are facing each other then a Patchy $(P)$ bond is formed with probability $\alpha_a$ and if a bond already exists it is broken with a probability $\beta_a$, such that $\frac{\alpha_a}{\alpha_a+\beta_a}=1-\exp(-u_a)$ see Fig. \ref{PNPI}b. When particles are in interaction range and the patches are not facing each other, they form Non Patchy Isotropic ($NPI$) bond with a probability $\frac{\alpha_i}{\alpha_i+\beta_i}=1-\exp(-u_i)$ \cite{prabhu2014brownian}, where $\alpha_i$ and $\beta_i$ are the probabilities to form and break the isotropic bond respectively see Fig. \ref{PNPI}a. The collection of these bonded particles is considered to be a cluster. In the present work we have considered $u_i$ ranging from 0 to 2.6 and $u_a$ values ranging from 0 to 10. 

\begin{figure}
\includegraphics[height=6cm,width=8.5cm]{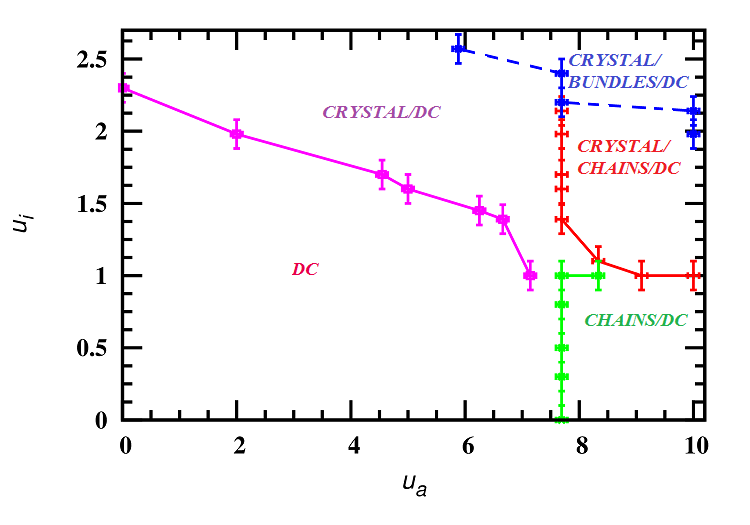}
  \caption{Calculated phase diagram in the $u_i-u_a$ plane. Different phases formed in different regions of phase diagram are shown. The dotted line shows the region of bundles which is a thermodynamically favored state.
    }
  \label{phasediagram}
\end{figure}

\begin{figure}
\includegraphics[height=8.5cm,width=9cm]{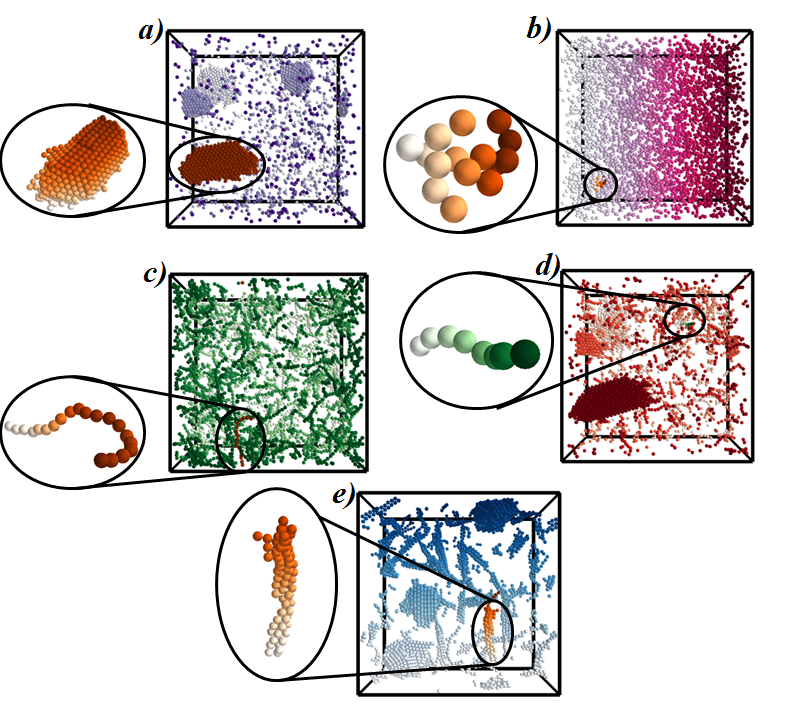}
  \caption{ Snapshot of different phases formed in different regions of phase diagram.}
  \label{snapshot}
\end{figure}

The phase diagram for two-patch model has been shown in Fig. \ref{phasediagram} where we observe the existence of five different phases. At intermediate to high isotropic ($1.39<u_i<2.57$) interaction and at low to intermediate anisotropic interaction ($2<u_a<6.67$) strength we observe the formation of crystals and $DC$ (disordered clusters) as shown in snapshot of the system in Fig. \ref{snapshot}a. As can be observed this region of phase diagram is dominated by isotropic part of the potential and the phase separation is similar to what is observed for system having only isotropic potential \cite{babu2006phase}. For the well width $\epsilon=0.1$, the isotropic square well system phase separates at $u_i\geq2.57$ \cite{babu2006phase} while in presence of anisotropic interaction we observe phase separation shifts to $u_i<2.57$. Below this region, as we decrease $u_i$ and $u_a$ we observe the formation of only $DC$ as shown in Fig. \ref{snapshot}b. In this region both the isotropic as well as anisotropic part of the potential is relatively small. When $u_a$ is increased ($u_a > $ 7.69) keeping the same $u_i=1.3$ we observe the formation of $DC$ as well as chains  see  Fig. \ref{snapshot}c. This region is dominated by anisotropic part of the potential as $u_a>u_i$. As mentioned before due to the small patch size, only one bond per patch is possible and as anisotropic bond strength is large, it leads to the formation of chains. Due to the presence of a weak isotropic interaction we also observe a small fraction of $DC$. When we increase $u_i$, for the same $u_a>7.69$ we observe the formation of crystals along with chains as isotropic part of the potential is playing a competing role (upto $u_i=2.14$) as shown in Fig. \ref{snapshot}d. As $u_i$ is further increased ($u_i>2.14$) crystals and bundles along with $DC$ are observed as shown in Fig. \ref{snapshot}e. We also observe that the finite bundle that is formed always has helical arrangement. It has already been shown that a bundle with an helical arrangement is a thermodynamically favored state \cite{grason2007chirality,yang2010self,turner2003twisted}, which is probably why we are observing the bundles with crystals \cite{Wellems,Makowski,Pumphrey,Jones}. For the case of single patch particle it was reported that finite bundles have a lower free energy compared to the lamellar structure which was the equilibrium structure \cite{preisler2013phase}.


\begin{figure}
\includegraphics[height=8cm,width=8.5cm]{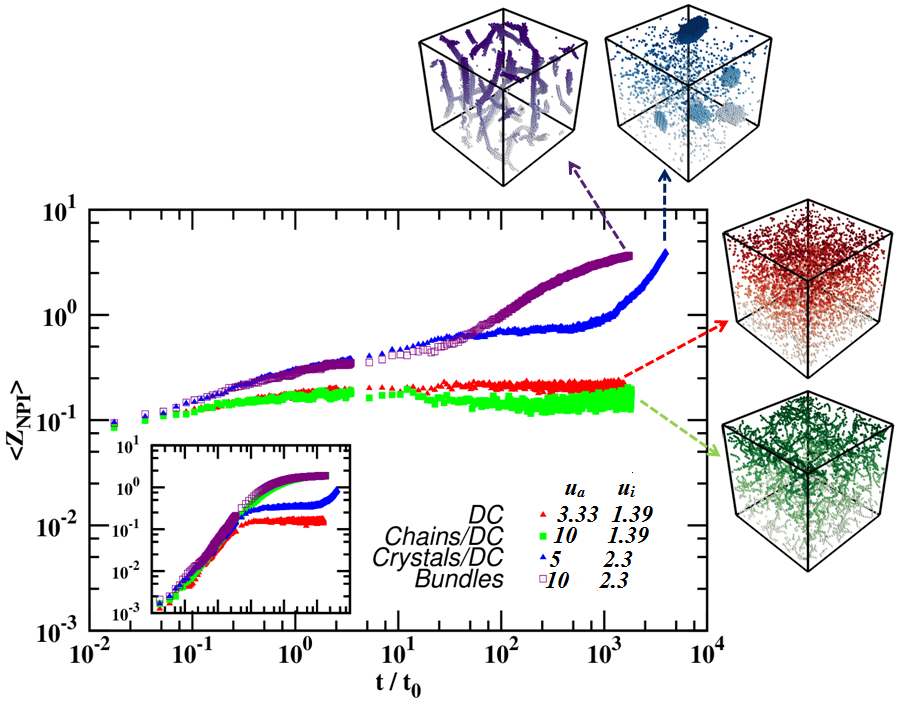}
  \caption{$<Z_{NPI}>$  is plotted with respect to reduced time starting from a random configuration of particles for different points in the phase diagram. $<Z_P>$ is plotted as a function of physical time in the inset. The visual image of the network formed by bundles is also shown as an inset.}
  \label{bundlechains}
\end{figure}

In order to understand the kinetics of formation of different phases, we follow the average number of neighbors bonded via only isotropic interaction $<Z_{NPI}>$ and only patchy interaction $<Z_P>$ (in the inset) as a function of reduced time starting from random distribution of particles.  For $u_i=1.39$ and $u_a=3.33$ we observe only disordered clusters where $<Z_P>$ and $<Z_{NPI}>$ increases very slowly and equilibrates around a value $10^{-1}$ as the isotropic and anisotropic potential are small. For chains$/DC$ $<Z_P>\sim 2$, showing that all the patches are bonded due to high anisotropic interaction strength ($u_a=10$) and $<Z_{NPI}>$ equilibrates  around $0.17$ due to low isotropic interaction strength $(u_i=1.39)$. When $u_i=2.3$ and $u_a=5$ we observe the formation of crystals as seen by the sudden upturn in the average number of neighbors $<Z_{NPI}>$, which is also observed in the case of $<Z_P>$. When the anisotropic part of the potential is increased $(u_a=10)$, we also  observe that $<Z_{NPI}>$ increases and a sudden upturn is observed similar to crystal structure formation, but $<Z_P> \sim 2$ indicates that we are observing chain like structure. A visual inspection of the images as shown in Fig. \ref{bundlechains} we observe a network made up of bundles similar to that seen in actin filaments \cite{sept2001thermodynamics}. 

\begin{figure}
\includegraphics[height=8.5cm,width=8.5cm]{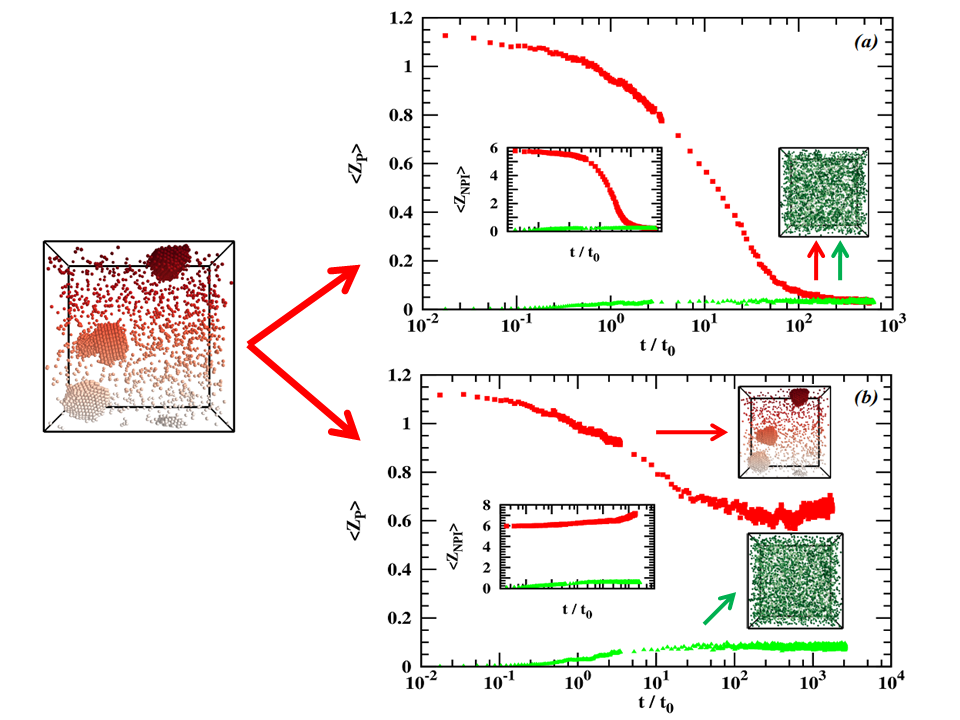}
  \caption{(a) $<Z_{P}>$ is plotted with respect to reduced time at $u_{i}=1.39$ and $u_a=3.03$ and inset shows the evolution of $Z_{NPI}$. (b) $<Z_P>$ is plotted at $u_{i}=2.14$ and $u_a=3.03$ and inset shows the evolution of $Z_{NPI}$. Triangles indicate the evolution of the system starting from random distribution of particles and squares indicate the evolution starting from a phase separated system formed at $u_i=2.14$ and $u_a=5.0$.}
  \label{cooling}
\end{figure}

Colloids with short range interaction phase separates into crystal-gas phase and this transition is first order in nature. To know the equilibrium states at different points $(u_a, u_i)$ on phase diagram, we have cooled the system from a random configuration to $(u_a,u_i)$ as well as heated an already phase separated system (crystallized) configuration to the same point in phase diagram.

In Fig. \ref{cooling}a we have plotted $<Z_{P}>$ as a function of reduced time, the squares indicate the evolution of system to $(3.03,1.39)$ starting from an already phase separated system formed at $(5,2.14)$ (see snapshot to the left of Fig. \ref{cooling}) and triangles indicate the evolution to $(3.03,1.39)$ starting from a random configuration of patchy particles. We observe that the dense crystal phase (indicated by squares) melts completely and meets with the curve starting from a random distribution indicating that the $DC$ formed is the equilibrium phase (see snapshot in the right side of Fig \ref{cooling}a). We also observe that similar trend is observed in the case of $<Z_{NPI}>$ as shown in the inset of Fig \ref{cooling}a. In Fig. \ref{cooling}b squares indicate the evolution of system to $(3.03,2.14)$ starting from a phase separated system formed at $(5,2.14)$ and triangles indicate the evolution of $Z_{P}$ to $(3.03,2.14)$ starting from a random distribution of particles. We observe that dense phase initially evaporates indicated by drop in $<Z_P>$ value, but crystal grows again indicating that system formed  is metastable in nature (see snapshot to the right top of Fig. \ref{cooling}b). In the inset we can observe that $<Z_{NPI}>$  of the dense phase increases continuously indicating that the crystals are not melting in this particular case. When we start the simulation from a random configuration and quench the system to $(3.03,2.14)$, the system is stuck in a meta-stable state and we observe only $DC$ (see snapshot to the right bottom of Fig. \ref{cooling}b). From a randomly distributed system when we quench the system to $(10,2.14)$ which is in the crystal/bundle region of the phase diagram we observe a network of bundles (see snapshot in Fig. \ref{bundlechains}). When we start from an already phase separated system formed in the crystal$/DC$  region $(5,2.14)$ we observe that the crystals in the system do not melt, whereas the $DC$ aggregate together to form finite helical bundles as shown in Fig. \ref{snapshot}e (also see Supplemental material \cite{Supplementary}).

\begin{figure}
\includegraphics[height=10cm,width=8.5cm]{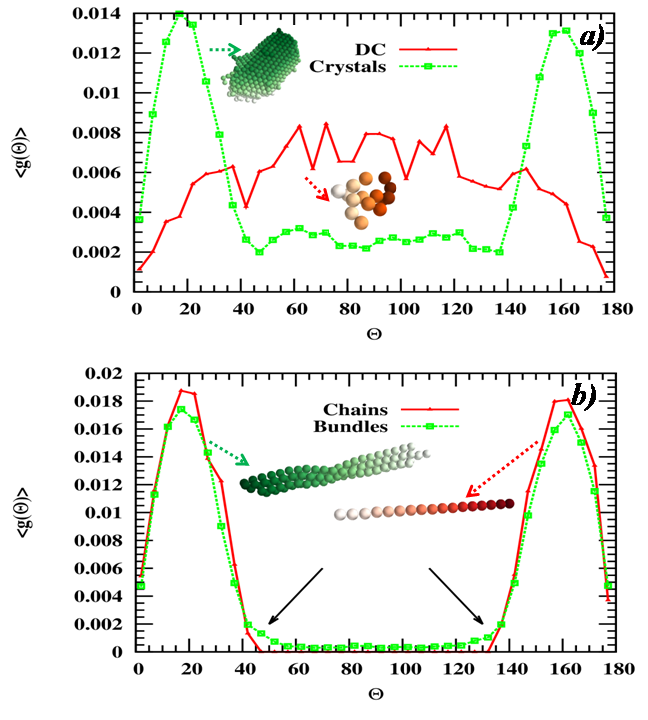}
  \caption{$g({\theta})$ as a function of $\theta$ is plotted for different phases of the phase diagram.}
  \label{gtheta}
\end{figure}
To differentiate between different structures we plot $g(\Theta)$ in Fig. \ref{gtheta}, which gives the probability of occurrence of angle $\Theta$ between patch vectors of adjacent particles \cite{malhotra2018aggregation}. In Fig. \ref{gtheta}a we have plotted the $g(\Theta)$ for the crystal structure and for the $DC$ cluster as obtained from the simulations. For the case of disordered cluster we observe that all the angles  between $20^{\circ}$ and $140^{\circ}$ are possible and hence these clusters are called as disordered cluster. We have isolated the crystals from the phase separated system and then calculated $g(\Theta)$ for the crystals alone. We observe $2$ prominent peaks for the crystal at $20^{\circ}$ and $160^{\circ}$, which indicates that the constituent spheres of the crystal are aligned along the patch vector. Along with this intermediate angles are also possible with a much smaller probability which may be due to the particles along the surface of the crystal. In Fig.\ref{gtheta}b we have plotted the $g(\Theta)$ after isolating the bundles and chains from a phase separated system. In this case the anisotropic interaction $u_a=10$ is greater than isotropic interaction $u_i=1.4$ and we observe chain formation. When the isotropic interaction is higher $u_i=2.3$ we observe the formation of bundles. The two peaks we observe are consistent with the fact that for a chain as well as a bundle the patch vector of the spheres is aligned either parallel or anti parallel with the neighboring sphere. It is already shown that once the system forms bundles it will twist into a helical structure and this is considered to be a thermodynamically favored state \cite{grason2007chirality,yang2010self}. Similar kind of structures  are observed in biological systems like sickle cells \cite{turner2003twisted} and actin filaments \cite{sept2001thermodynamics}. Due to the finite twist as predicted by Turner \textit{et al.} \cite{turner2003twisted} we are observing only a small tail in the distribution of $g(\Theta)$ at $40^{\circ}$ and $130^{\circ}$ for the parallel and anti parallel patch vectors as shown by arrows in Fig. \ref{gtheta}b. Due to the presence of this twist we can easily differentiate between a bundle and a chain apart from the visual confirmation.

In the present work, the formation of various phases such as disordered clusters, chains+crystals and crystals+ thermodynamically favored bundles is achieved by tuning the relative strength of $u_i$ and $u_a$. The kinetics of formation of helical bundles in our system is distinct from formation of helices as proposed by Chen \textit{et al.}\cite{chen}. In their work they studied the helical bundle formation of Janus particles while in the present work we have studied particles with $2$ patch along with isotropic interaction. Chen \textit{et al.}\cite{chen} proposed step-by-step addition of individual particles forming small clusters which later fuse together into fibrillar triple helices of finite diameter. While in the present study only when we have relatively high isotropic and anisotropic interaction we observe the presence of thermodynamically favored bundles. We have also shown that by tuning $u_a$ and $u_i$  we were able to mimic the structures observed for glucose isomerase protein \cite{van2018molecular}, which include nanorods (chains), crystals, disordered clusters and bundles. The bundles which we observe seems to be helical in structure which is a thermodynamically favored state as reported by Grason \textit{et al.} \cite{grason2007chirality,yang2010self},  which is similar to that observed in sickle  hemoglobin fibres \cite{turner2003twisted}. It will be interesting to study the kinetics as well as the structural properties by varying patch sizes. 
\\We thank HPC padum and Badal of IIT Delhi for providing us the necessary computational resources.

\end{document}